\documentstyle[11pt]{article}
\def\th{\theta} 
\def\pa{\partial}
\def\k{\kappa}
\def\g{\gamma} \def\G{\Gamma} %\mbox{\boldmath $A$}
\def\a{\alpha}
\def\b{\beta}
\def\d{\delta}

\def\k{\kappa}
\def\l{\lambda} \def\L{\Lambda}
\def\m{\mu}
\def\n{\nu}

\def\s{\sigma} 
\def\t{\tau}
 
\def\o{\omega} 
\def\mn{{\mu\nu}}
\def\ab{{\alpha\beta}}
\def\be{\begin{equation}}
\def\ee{\end{equation}}
\def\bea{\begin{eqnarray}}
\def\eea{\end{eqnarray}}
\begin{document}

\begin{center}
{\large\bf Self-Interaction and Gauge Invariance}

S. Deser\thanks{Supported by USAF OAR under Grant AFOSR 70-1864.}

Physics Department, Brandeis University, Waltham, MA\\ and
Nordita, Copenhagen

Received: 17 October 1969
\end{center}

\begin{center}
\parbox[b]{4in}{\footnotesize This article, the first paper in Gen.\
Rel.\ Grav.\ [{\bf 1} (1970) 9], is now somewhat inaccessible; the
present posting is the original version, with a few subsequent
references appended.}
\end{center}

\abstract{A simple unified closed form derivation of the
non-linearities of the Einstein, Yang-Mills and spinless ({\it
e.g.}, chiral) meson systems is given. For the first two, the
non-linearities are required by locality and consistency; in all
cases, they are determined by the conserved currents associated
with the initial (linear) gauge invariance of the first kind. Use
of first-order formalism leads uniformly to a simple cubic
self-interaction.}
 \setlength{\parskip}{.1in}

\noindent{\large\bf Introduction}

The Maxwell and Einstein fields are, respectively, the most and
least linear of gauge theories. The electrical neutrality of the
photon reflects the absence of self-interaction, while at the
other extreme, the gravitational field equations are an infinite
series in the metric due to the gravitational `weight' of
gravitons and of their interaction energy. Between these extremes
stand theories with internal gauge symmetry, typified by the (spin
1) Yang-Mills field and by (spin 0) chiral Lagrangians. We wish to
give a simple physical derivation of the non-linearity of these
theories, using a now familiar argument ({\it e.g.}, [1--6])
leading from the linear massless spin 2 field to the full Einstein
equations. This argument, which stresses the self-interaction
(rather than gauge invariance) aspects, proceeds by adjoining to
the initially linear theory a source which is obtained from the
free part itself, a further source due to this one, etc., thus
introducing new, non-linear terms in the action. The various
non-linearities are thereby exhibited as specific
self-interactions. We shall present a unified derivation, based on
use of first-order actions, of the non-linearities of the above
fields. All of them will emerge precisely as having cubic
Lagrangians of the same generic form. In particular, the Einstein
equations will be derived in one (closed form) step, rather than
as an infinite series. There, consistency implies universal
(including self-) coupling, and therefore the equivalence
principle.

\noindent{\large\bf Metric Field}

The Einstein equations may be derived non-geometrically [1--6] by
noting that the free massless spin 2 field equations,
\bea%1
\lefteqn{R^L_\mn (\phi ) -  \textstyle{\frac{1}{2}} \: R^L_{\a\a}
(\phi
) \eta_\mn \equiv G^L_\mn (\phi )} \\
&&\equiv  [(\eta_{\m\a} \eta_{\n\b} - \eta_{\m\n} \eta_{\a\b})
\Box + \eta_\mn \pa^2_{\a\b} + \eta_{\a\b} \pa^2_\mn - \eta_{\m\a}
\pa^2_{\n\b} - \eta_{\m\b} \pa^2_{\n\a}] \phi_{\a\b} = 0 \nonumber
 \eea
whose source is the matter stress-tensor $T_\mn$, must actually be
coupled to the {\it total} stress-tensor including that of the
$\phi$-field itself. That is, while the free-field equations (1)
are of course quite consistent as they stand, this is no longer
the case when there is a dynamical system's  $T_\mn$ as a source.
For then the left side, which is identically divergenceless, is
inconsistent with the right, since the coupling implies that
$T^\mn_{~~,\n}$ as computed from the matter equations of motion,
is no longer conserved. To remedy this,\footnote{Consistency and
linearity may also be reconciled if locality is abandoned [14].}
the stress tensor $^2\th_\mn$ arising from the quadratic
Lagrangian $^2L$ responsible for equation (1) is then inserted on
the right. But the Lagrangian $^3L$ leading to these modified
equations is then cubic, and itself contributes a cubic
$^3\th_\mn$. This series continues indefinitely, and sums (if
properly derived!) to the full non-linear Einstein equations,
$G_\mn (\eta_\ab + \phi_\ab ) = -\k T_\mn$, which are an infinite
series in the deviation $\phi_\mn$ of the metric $g_\mn$ from its
Minkowskian value $\eta_\mn$. Once the iteration is begun (whether
or not a $T_\mn$ is actually present), it must be continued to all
orders, since conservation only holds for the full series
$\sum^\infty_2 \, ^n\th_\mn$. Thus, the theory is either left in
its (physically irrelevant) free linear form (1), or it {\it must}
be an infinite series. The actual process of inserting the
$\th_\mn$ of the system at each step is the prototype of  our
method: the `current' on the right is that generated by the
initial constant gauge invariance of the theory. In this case, the
$\th_\mn$ are the coefficients of local Lorentz transformations,
since the invariance is that under rigid Lorentz rotations. This
procedure is necessary when—-and only when-—there is also an
initial gauge invariance of the second kind (here $\d\phi_\mn =
\pa_\m \xi_\n + \pa_\n \xi_\m$) which implies identical
conservation of the free field part, although this invariance is
in fact violated by the iteration procedure. The current is
defined at each step by invariance under constant transformations.

We now derive the full Einstein equations, on the basis of the
same self-coupling requirement, but with the advantages that the
full theory emerges in closed form with just one added (cubic)
term, rather than as an infinite series, and that no special
`gauge' such as $g^\mn_{,\n}=0$ need be introduced. This is made
possible by use of first-order form, in which the metric and
affinity are {\it a priori} independent, and by taking as initial
variables the linearizations not of ${\rm g}_\mn$, but of $g^\mn$,
the contravariant metric density,

We begin by recording the full Einstein action in first-order form
\bea%2
I & \equiv & \int \, d^4x \, {\cal R} \equiv \int \, d^4x \, g^\mn
\,
R_\mn \\
& = & \int d^4x \, g^\mn [\G^\a_{\mn,a} - \G^\a_{\m\a,\n} +
\G^\a_{\n\m}\G^\b_{\ab} - \G^\a_{\b\m}\G^\b_{\a\n} ] \nonumber
 \eea
which yields the field equations
$$%3a
\G^\a_\mn = \Bigg\{ \!\!\! \begin{array}{c} \a \\ \mn \end{array}
\!\!\! \Bigg\} \eqno{(3{\rm a})}
$$
$$%3b
R_\mn \equiv \G^\a_{\mn ,\a} - \textstyle{\frac{1}{2}} \,
\G^\a_{\a\m ,\n} - \textstyle{\frac{1}{2}}\, \G^\a_{\a\n ,\m} +
\G^\a_{\mn}\G^\b_{\ab} - \G^\a_{\b\m}\G^\b_{\a\n} = 0 \; .
\eqno{(3{\rm b})}
$$
\renewcommand{\theequation}{\arabic{equation}}
\setcounter{equation}{3}
 Here $\left\{ \! ~^\a_\mn \! \right\}$ is the
Christoffel symbol constructed from the metric, and $\G^\a_\mn$
and $g^\mn$  have been varied independently. Note that the action
is just cubic in these basic variables. The free massless spin 2
theory (linearized approximation) may be represented by the
quadratic action
\be%4
I^L \equiv \int \, d^4x \, {\cal R}^Q = \int \, d^4x [h^\mn
(\G^\a_{\mn ,\a} - \G_{\m , \n}) + \eta^\mn (\G^\a_\mn \G_\a -
\G^\a_{\b\m} \G^\b_{\a\n})] \ee
 with field equations
$$%5a
2\G^\a_\mn - \eta^a_{\;\;\m} \G_\n - \eta^\a_{\;\;\n} \G_\m =
h^\mn_{\;\;\;{,\a}} - h^{\n\a}_{\;\;\;{,\m}} -
\textstyle{\frac{1}{2}}\, \eta_\mn h^\b_{\b,\a}\eqno{(5{\rm a})}
$$
$$%5b
\G^\a_{\mn ,\a}-\textstyle{\frac{1}{2}}\,\G_{\m ,\n}
-\textstyle{\frac{1}{2}}\,\G_{\n ,\m} = 0 \eqno{(5{\rm b})}
$$
where $\G_\m \equiv \G^\a_{\m\a}$. It differs from (2) only in the
replacement of $g^\mn$ by $\eta^\mn$ in the cubic term. As all
indices are moved by $\eta_\mn$, we need only keep track of the
symmetry of $h^\mn$ and of the bottom indices of $\G^\a_\mn$.
Differentiation of (5a) with respect to $\a$ yields the linear
equation
\renewcommand{\theequation}{\arabic{equation}}
\setcounter{equation}{5}
\be%6
2R^L_\mn \equiv \Box h^\mn - h^{\m\a}_{\;\;\;{,\a\n}} -
h^{\n\a}_{\;\;\;{,\a\m}} - \textstyle{\frac{1}{2}}\, \eta_\mn \Box
h_{\a\a} = 0
 \ee
which are equivalent to (1), with the relation $\phi_\mn = - h_\mn
+ \frac{1}{2}\, \eta_\mn h_{\a\a}$. We now demand that equation
(6) be augmented by the source:\footnote{It is equivalent and
saves computation to work with $R_\mn$ instead of $G_\mn$ which is
why the source is $\t_\mn \equiv \d I/\d\psi^\mn$. Also, we set
the proportionality constant $\k$ between $R$ and $\t$ to unity,
since it gets reabsorbed anyhow in the final redefinition $g^\mn =
\eta^\mn + \k h^\mn$. Of course, $\k$ will appear in front of the
matter stress tensor.}  $\t_\mn \equiv T_\mn - \frac{1}{2} \,
\eta_\mn T_{\a\a}$, where $T_\mn$ is the stress-tensor of the
linear action of equation (4). It is very simply computed in the
usual (Rosenfeld) way as the variational derivative of $I^L$ with
respect to an auxiliary contravariant metric density $\psi^\mn$,
upon writing $I^L$ in `generally covariant form', $I^L (\eta
\rightarrow \psi )$, with respect to this metric. Note that this
does not presuppose any geometrical notions, being merely a
mathematical shortcut in finding the symmetric stress-tensor of
$I^L$. We could also obtain it by the (equivalent) (Belinfante)
prescription of introducing local Lorentz transformations. The
covariant action is simply (4) with $\eta^\mn \rightarrow
\psi^\mn$ and $\psi$-covariant derivatives in the $h\pa\G$ term:
\bea%7
\d I^L(\psi ) & \equiv & \int d^4x \, \d\psi^\mn [(h^\ab \G_\l -
2h^{\rho\b} \G^\a_{\rho\l} + h^{\rho\t}  \G^\a_{\rho\t} \,
\d^\b_\l ) (\d
C^\l_{\a\b}/\d\psi^\mn ) \nonumber \\
& + & (\G^\a_\mn \G_\a - \G^\a_{\b\m} \G^\b_{\a\n})]
 \eea
where $C^\l_\ab$ is the Christoffel symbol of the $\psi$. We have
chosen to let $h^\mn$ transform as a contravariant tensor density
and $\G^\a_\mn$ as a tensor in this auxiliary space. Since we are
only interested in getting $\d I/\d\psi^\mn$ at $\psi = \eta$, it
is straightforward to vary $C^\l_\ab$, keeping only the linear
terms $\sim \pa\psi$, to obtain
 \bea%8
\t_\mn & \equiv & \d I^L/\d\psi^\mn = (\G_\a\G^\a_\mn -
\G^\a_{\b\m} \G^\b_{\a\n} ) - \s_\mn \nonumber \\
2\s_\mn & \equiv & \pa_\a [ \eta_\mn ( h^{\l\rho} \G^\rho_{\l\a} -
\textstyle{\frac{1}{2}} \, h^{\l\l} \, \G_\a ) + (h^\mn \, \G_\a -
h^{\m\a} \G_\n - h^{\n\a} \G_\m ) \nonumber \\
&& + \; h^\ab (\G^\m_{\b\n} + \G^\n_{\b\m}) + h^{\m\rho}
(\G^\a_{\rho\n} - \G^\n_{\a\rho} ) + h^{\n\rho} (\G^\a_{\rho\m} -
\G^\m_{\a\rho})] \; .
 \eea
We now assert that the action which leads to the desired equation
$R^L_\mn = -\t_\mn$ is
\be%9
I = I^L + \int d^4x \, h^\mn (\G^\a_\mn \G_\a - \G^\a_{\b\m}
\G^\b_{\a\n}) \; .
 \ee
Note that we have {\it not} added the full $h^\mn\t_\mn$, but
rather used the simple part of $\t_\mn$. We also note that, if our
assertion is correct, {\it no iteration will be needed} as the
cubic term in (9) is in fact $\psi$-independent since $h^\mn$ is a
density.   Thus $\d I/\d\psi^\mn = \d I^L/\d \psi^\mn$, and (9)
constitutes the full theory as it must, since it is precisely the
Einstein action (2) with  the identification $g^\mn = \eta^\mn +
h^\mn$.  To check that (9) is correct,  we compute $R^L_\mn$ from
the field equations (3).  These differ from the linear ones in two
respects. The first is the $(\G\G -\G\G )$ term in (3b), which is
the simple part of our $\t_\mn$.   The second is that $\G$ is now
the full Christoffel symbol, {\it i.e}, that (3a)
reads\footnote{The usual infinite non-linearity of the Einstein
equations appears when (5a) is solved for $\G$ which involves the
matrix inverse of $(h^\mn + \eta^\mn )$.  Note that $\eta^\mn$
assures the existence of this inverse at infinity, where $h^\mn$
is assumed (like any other field) to vanish.}
\be%10
-h^\mn_{\;\;\;{,\a}} + (\eta^\mn + h^\mn ) \G_\a - (\eta^{\m\rho}
+ h^{\m\rho})\G^\n_{\a\rho} - (\eta^{\n\rho} +
h^{\n\rho})\G^\m_{\a\rho} = 0
 \ee
and contains bilinear $h\G$ terms, unlike (5a). If we
differentiate (10) with respect to $\a$ after cycling on the
indices and separate the linear and quadratic terms we find
precisely
\be%11
2\G^\a_{\mn ,\a} - \G_{\m , \n} - \G_{\n , \m} = 2R^L_\mn -
2\s_\mn
 \ee
so that with (3b), the net result is just the desired one,
\be%12
R^L_\mn = - \t_\mn \; .
 \ee
Thus, thanks to use of first-order form, together with the use of
the natural variable $g^\mn$, in terms of which the full Einstein
action is cubic, the derivation involved just one direct step.
Note also that the two parts of $\t_\mn$ have a different origin
both in the field equations and in the $\d /\d\psi$ procedure. The
$\G\G$ term is the direct part of $\d I/\d\psi$ and of the obvious
quadratic contribution in the $\pa\G$ equation due to the cubic
term.  The $\s_\mn$ part is more subtle: it is only for spin $>$1
that the kinetic `$p\dot{q}$' term is unavoidably $\psi$-dependent
and so contributes to the stress tensor.  Thus for spin 1, the
corresponding term is $F^\mn (\pa_\m A_\n - \pa_\n A_\m )$, which
is covariant as it stands, taking $F^\mn$ to be a density, and
likewise we have the covariant form $\pi^\m \pa_\m\phi$ for spin
0, with $\pi^\m$ a density. However, it is well known that higher
rank tensors {\it e.g.}, symmetric second-rank ones must have
explicit covariant derivatives.  Likewise the $\s$ contribution in
the field equations is due to the non-linearity  of the $\G - h$
relation, so that it arises from the difference between $\G$ and
its linear part (in $h$).

Finally, we return to the coupling of matter.  The matter source
is taken initially to be the conserved current associated with
invariance of the free matter system under rigid Lorentz
transformations, namely $T^M_\mn (\eta )$, and does not, at this
stage, depend on $h^\mn$.  It is easy to show that the correct
coupling is according to the usual minimal prescription $I^M
(\eta^\mn ) \rightarrow I^M (\eta^\mn + h^\mn )$:  For, on the one
hand, the right side of the Einstein equation (12) is to be
$\t^M_\mn \equiv \d I_M (\eta +h )/\d\psi^\mn$ at $\psi = 0$,
namely the total matter stress tensor. On the other, viewed as an
Euler-Lagrange equation, (12) is effectively $\d I^{\rm Tot}/\d
h^\mn = 0$. Thus, we must have $\d I^M (\psi )/\d\psi |_0 = \d I^M
(h)/\d h$ whose solution is clearly $I^M = I^M (\eta^\mn + h^\mn
)$, remembering that $\eta \rightarrow \eta + \psi$ .

Consistency has therefore led us to universal coupling, which
implies the equivalence principle. It is at this point that the
geometrical interpretation of general relativity arises, since
{\it all matter} now moves in an effective Riemann space of metric
$g^\mn \equiv \eta^\mn + h^\mn$, and so the initial flat
`background' space $\eta^\mn$ is no longer observable.

\noindent{\large\bf Yang-Mills Field}

Consider now as an example of vector theories with internal
symmetry, the Yang-Mills field, with $SU_2$ invariance [7]. We
begin with the linear system, a triplet of free massless vector
fields with potentials $A^a_\m$ and field strengths $F^a_\mn$,
where $a$=1, 2, 3 is the internal index. The first-order action
 $$%13a
I_0 = - \textstyle{\frac{1}{2}} \, \int \, d^4 x [{\bf F}_\mn
\cdot (\pa_\m {\bf A}_\m - \pa_\n{\bf A}_\m )
-\textstyle{\frac{1}{2}} \,{\bf F}_\mn \cdot {\bf F}_\mn ]
\eqno{(13{\rm a})}
$$
yields the field equations
$$%13b
\pa_\m {\bf F}_\mn = 0 \; , \;\;\;\;\; {\bf F}_\mn = \pa_\m {\bf
A}_\n - \pa_\n {\bf A}_\m \eqno{(13{\rm b})}
$$
upon independent variation of ${\bf A}_\m$ and ${\bf F}_\mn$ (we
use vector notation for the isotopic index throughout). This set
of free Abelian gauge fields is invariant under the usual Maxwell
transformations of the second kind, $ {\bf A}_\m \rightarrow {\bf
A} + \pa_\m \mbox{\boldmath $\L$}, \; {\bf F} \rightarrow {\bf F}$
which imply that $\pa_\m{\bf F}^\mn$  is identically conserved.
This property will require self-coupling (for consistency where
sources are present). Its form is determined by the invariance
under constant internal rotations\footnote{If we used the current
corresponding to rotations about a particular direction only, the
resulting theory would actually be inconsistent [8], so that the
full symmetry must be exploited.}
\renewcommand{\theequation}{\arabic{equation}}
\setcounter{equation}{13}
\be%14
\mbox{\boldmath $\th$}\rightarrow\mbox{\boldmath $\th$} +
\mbox{\boldmath $\th$}\! \times \!\mbox{\boldmath $\o$}
 \ee
where $\mbox{\boldmath $\th$}$ stands for ${\bf A}$ or ${\bf F}$.
(The absence of such an invariance for the single real Maxwell
field is responsible for its linearity.) The associated conserved
current is
\be%15
{\bf j}_\m (\chi ) \equiv \d I/\d \pa_\m \mbox{\boldmath $\o$} (x
) = g{\bf F}_\mn \times {\bf A}_\n
 \ee
where the variable gauge transformations $\mbox{\boldmath $\o$} (x
)$ have just been introduced as a convenient means of obtaining
the current according to the usual Noether theorem argument. If we
now augment the quadratic action $I_0$ with the self-coupling term
${\bf j}_\m\cdot{\bf A}_\m$, which retains (constant) rotation
invariance, we have
\be%16
I = I_0 + \int \, d^4x \; {\bf j}_\m\cdot {\bf A}_\m = I_0 +
\textstyle{\frac{1}{2}} \, g \int \, d^4x \,{\bf A}_\m \cdot{\bf
F}_\mn \times{\bf A}_\n
 \ee
with field equations
 \bea%17
{\bf F}_\mn & = & \pa_\m{\bf A}_\n - \pa_\n{\bf A}_\m + g{\bf
A}_\m \times
{\bf A}_\n \nonumber \\
\pa_\n{\bf F}_\mn & = & +g {\bf F}_\mn \times{\bf A}_\n = {\bf
j}_\m
 \eea
As in our treatment of the general relativistic case, addition of
the same generic cubic term in (16) yields the full theory without
further iteration. This may be seen in two different ways: the
self-interaction ${\bf A}\cdot {\bf F} \!\times \!{\bf A}$ does
not involve explicit derivatives and hence will not contribute to
a further ${\bf j}_\m$ term, as defined by (15). Alternately, the
${\bf j}_\m$ defined in (15) is already conserved as a consequence
of the {\it full} equations (17), which are thus consistent as
they stand:
\be%18
\pa_\m{\bf j}_\m = g({\bf F}_{\mn ,\m}\!\! \times \!\! {\bf A}_\n
+ {\bf F}_\mn \!\! \times \!\! {\bf A}_{\n ,\m} ) = 0 \; .
 \ee
The action (16) is, of course, the complete Yang-Mills action in
first-order form. It is invariant under an extended group of gauge
transformations of the second kind, although this was not required
initially (as was the case for the corresponding general
coordinate invariance of the full Einstein equations). This is a
basic difference between the present and those derivations [9]
which are based on the extended invariance requirements.

The above derivation exhibited the Yang-Mills theory as one in
which the isotopic current is the source of $\pa_\m {\bf F}_\mn$,
rather than of the linear expression $\pa_\m (\pa_\m {\bf A}_\n
-\pa_\n {\bf A}_\m )$. Indeed, the two differ by a term $\pa_\m
({\bf A}_\m \times {\bf A}_\n )$ which is identically conserved,
but not obtained from gauge invariance. This is just the converse
of the Einstein situation, where $\t_\mn$ was the source of the
$R^L_\mn (h)$ rather than of $\G^\a_{\mn , a} - \G_{\m ,\n}$
(which is not identically conserved).

Had we started from the second-order formalism, so that
\be%19
I_0 = - \textstyle{\frac{1}{4}} \int  d^4x (\pa_\m {\bf A}_\n -
\pa_\n {\bf A}_\m )^2
 \ee
its invariance under $\d{\bf A} = {\bf A} \times \mbox{\boldmath
$\o$}$ yields the current
\be%20
^1{\bf j}_\m = g{\bf A}_\n \times (\pa_\m{\bf A}_\n - \pa_\n{\bf
A}_\m ) \; .
 \ee
We would then expect to make the addition
 \be%21
 I_1 =\textstyle{\frac{1}{2}} \int  d^4x \: ^1{\bf j} \cdot{\bf A}
=\textstyle{\frac{1}{2}} \, g \int d^4x\,{\bf A}_\m \cdot
(\pa_\m{\bf A}_\n - \pa_\n{\bf A}_\m ) \times {\bf A}_\n \; .
 \ee
 Since equation (21) still involves explicit derivatives, it would yield one
further iteration
 \be%22
 ^2 {\bf j}_\m = g^2 ({\bf A}_\m  \times {\bf A}_\n ) \times {\bf
 A}_\n\; ,\;\;\;\; I_2 =\textstyle{\frac{1}{4}} \, g^2 \, \int
 d^4x \,({\bf A}_\m \times {\bf A}_\n )^2
 \ee
 in which there are no derivatives left. The total action is the familiar
 second-order form of Yang-Mills theory,
\be%23
I = I_0 + I_1 + I_2 = - \,\textstyle{\frac{1}{4}} \, \int \, d^4x
\, {\bf F} \cdot {\bf F}\; , \;\;\;\; {\bf F}_\mn \equiv {\bf
D}_\m {\bf A}_\n - {\bf D}_\n {\bf A}_\m
 \ee
in terms of the ``$\frac{1}{2}$-covariant" derivative ${\bf D}_\m
\equiv (\pa_\m - \frac{1}{2} \, g {\bf A}_\m \times )$.

The final cubic (16) and quadratic (23) forms are of course
equivalent. However, the procedure leading to the latter does not
fulfill the original self-coupling postulate, since variation of
the first iteration (21) does not yield field equations with
$^1{\bf j}_\m$ as source, but has the additional $\pa_\n ({\bf
A}_\m \!\!\times \!\!{\bf A}_\n )$ term mentioned earlier.

Just as in the Einstein case, conservation of the current
permitted, but did not require, self-interaction in the absence of
sources. However, just as in the Einstein case, it {\it is}
necessary in the only interesting situation, in which a dynamical
current $J_\m$ interacts with the field. For example for a fermion
field $\mbox{\boldmath{$\psi$}}$, its current ${\bf J} \equiv
g\bar{\psi}\g)\pa_\m \mbox{\boldmath{$\t$}}\psi$ will {\it not} be
conserved as a result of the Dirac equation, but will obey a
`covariant conservation' law, and so cannot be consistently
coupled to the {\it linear} theory (since $\pa^2_\mn {\bf F}^\mn
\equiv 0$) even though ${\bf J}\cdot{\bf A}$ is rotationally
invariant. It then becomes necessary to introduce
self-interaction, that is transversality of the field equations
with respect to {\it covariant} differentiation $({\bf D}_\m{\bf
D}_\n{\bf F}^\mn = 0)$. Our argument is of  course no longer
compelling for a massive vector field, since the mass term can
always absorb the non-conserved part of the current without need
for non-linear terms, according to $M^2{\bf A}^\n\,_{,\n} = {\bf
J}^\n\,_{,\n} \sim {\bf J}_\n \times{\bf A}_\n$. However, it is
still  perfectly consistent to iterate and obtain the massive
version of Yang-Mills theory.

\noindent{\large\bf Spin Zero Systems}

Unlike the situation for tensor and vector fields, a non-linear
theory is not mandatory for spinless particles, because as we
shall see, there is no clash between external current
non-conservation and the free field equations.  However, it is
still possible to carry out the same procedure, and insist that
the source of the field be the {\it total} current, including that
of the massless field itself. We consider here one
example,\footnote{The simplest spin zero example, that of `scalar
gravitation', where a single scalar field is coupled to the trace
of its stress tensor, is treated in second-order form in [13] and
[10]. A later communication will deal with the Nordstr\"{o}m
theory.} which leads to the Sugawara theory [11,12] of currents.

Rather than obtain the chiral Lagrangian in one or another
particular spin zero representation, which would correspond to
deriving the non-linearities of Einstein theory in a particular
gauge, we shall reach it in a general form. To this end, consider
the quadratic action
 \be%24
I_0 = -\, \textstyle{\frac{1}{2}} \, \int \, d^4x [{\bf F}_\mn
\cdot (\pa_\m {\bf A}_\n - \pa_\n {\bf A}_\m ) + c {\bf A}_\m^2 ]
 \ee
which describes a triplet of purely longitudinal free massless
fields with field equations
$$%25a
\pa_\m {\bf A}_\n - \pa_\n {\bf A}_\m = 0 \eqno{(25{\rm a})}
$$
$$%25b
\pa_\m {\bf A}_\m  = 0  \; . \eqno{(25{\rm b})}
$$
This is the Abelian limit of Sugawara theory and is equivalent to
a triplet of free massless scalar fields [since (25a) implies
${\bf A}_\m = \pa_\m \mbox{\boldmath{$\phi$}}$, and (25b) yields
$\Box \mbox{\boldmath{$\phi$}} =0$]. If we now adjoin the current
term, which is just $I_1 \sim \int A\cdot F \times A$, as in the
Yang--Mills case (since the invariance is the same), we obtain
\renewcommand{\theequation}{\arabic{equation}}
\setcounter{equation}{25}
\be%26
I = I_0 + I_1 = - \, \textstyle{\frac{1}{2}} \, \int \, d^4x [{\bf
F}_\mn \cdot ({\bf D}_\m {\bf A}_\n - {\bf D}_\n {\bf A}_\m ) +
c{\bf A}^2_\m ]
 \ee
 where ${\bf D}_\m \equiv \pa_\m - \frac{1}{2}\, g{\bf A}_\m
 \times$.

This action has been discussed elsewhere [13], and shown to be a
Lagrangian formulation of the Sugawara model (extension to the
$SU_2 \times SU_2$ case is immediate). This is clear from the fact
that the resulting field equations are the usual
$$%27a
D_\m{\bf A}_\n - D_\n{\bf A}_\m \equiv \pa_\m {\bf A}_\n - \pa_\n
{\bf A}_\m - g {\bf A}_\m \times {\bf A}_\n = 0  \eqno{(27{\rm
a})}
 $$
$$%27b
\pa_\m {\bf A}^\m = 0  \eqno{(27{\rm b})}
 $$
and that the equal time commutation relations are also identical.
The action is again only cubic, although it becomes an infinite
series when expressed in terms of the spin zero pion field in one
or another of the non-linear realizations of the field equations,
corresponding to the choice of basic pion field $\pi$ to represent
the solution of (27a). The same procedure could be attempted
starting with the scalar representation
\renewcommand{\theequation}{\arabic{equation}}
\setcounter{equation}{27}
 \be%28
 I_0 = - \int \, d^4x [\mbox{\boldmath{$\pi$}}^\m \cdot
 \mbox{\boldmath{$\phi$}}_\m -  \textstyle{\frac{1}{2}}\,
 \mbox{\boldmath{$\pi$}}^2 ]
 \ee
of a triplet of massless pions, and using the invariance under $\d
\mbox{\boldmath{$\pi$}} = \mbox{\boldmath{$\pi$}} \times
\mbox{\boldmath{$\o$}}\, , \; \d  \mbox{\boldmath{$\phi$}} =
\mbox{\boldmath{$\phi$}} \times  \mbox{\boldmath{$\o$}} +
\mbox{\boldmath{$\o$}}^\prime$ [App.\ B]. However, this is
considerably more complicated than the above
representation-independent treatment and the self-coupling
prescription is not directly fulfilled at each step. Note that
there is no necessity here of adding the non-linear term, since
the original field equations in the presence of an external
current ${\bf J}_\m$ (coupling ${\bf J}\cdot{\bf A}$),
 \be%29
 0 = \pa_\m {\bf A}_\n - \pa_\n {\bf A}_\m \; , \;\;\;\; \pa_\m
 {\bf F}^\mn = c {\bf A}_\n + {\bf J}_\n
 \ee
allow for non-conserved ${\bf J}_{\n ,}$ with $\pa {\bf A} \sim
\Box \mbox{\boldmath{$\phi$}} \sim \pa {\bf J}$, as for massive
vector theory. Unlike the transverse Yang-Mills field, this purely
longitudinal gauge field does not possess the initial gauge
invariance ${\bf A} \rightarrow {\bf A} + \pa
\mbox{\boldmath{$\L$}}$ which  required the non-linearity there.
It is, however, natural to add the cubic  term so that covariant
derivatives enter throughout.

\noindent{\large\bf Appendix A}

\renewcommand{\theequation}{A.\arabic{equation}}
\setcounter{equation}{0}

We sketch here an amusing `geometrical derivation' of the Einstein
or Yang-Mills equations which is, however, less compelling than
that in text.  Consider the free spin 2 equations in terms of the
variables $\g^\m_\ab , \phi_\mn$ (where $\phi_\mn$ is ultimately
$g_\mn - \eta_\mn$). Then the field equations read
 \bea%A.1
 2\g^\a_{\mn ,\a}  -  \g_{\m ,\n} - \g_{\n ,\m} &=& 0 \nonumber \\
\g^\a_\mn & = &  \textstyle{\frac{1}{2}}\, \eta^\ab [\phi_{\b\m
,\n} + \phi_{\b\n ,\m} - \phi_{\mn ,\b} ] \; .
 \eea
In a background metric space they would then have the same form,
but with all derivatives replaced by covariant ones with respect
to the background, and with $\eta \rightarrow g$. We now {\it
identify} $\g$ and $\phi$ as small variations of the background
$\G$ and $g$. Then these covariant equations `integrate' to the
usual Einstein ones (3), using the Palatini identity
 $$
 \d R_\mn = (\d \G^\a_\mn )_{;\a} - \textstyle{\frac{1}{2}} (\d
 \G_\m )_{;\n} - \textstyle{\frac{1}{2}} (\d
 \G_\n )_{;\m}
 $$
together with the obvious one for $\d\G^\a_\mn$ and recalling that
$\d\G$, unlike $\G$ itself, is a tensor. In particular, (A.1) thus
represents the small oscillations of the Einstein field near flat
space or with respect to a local inertial frame where $\G=0$ and
$g = \eta$. Likewise we could start with the `flat space'
equations (13b) in terms of small oscillations ${\bf f}_\mn \, ,
\; {\bf a}_\n$ which, in the presence of an external ${\bf A}_\m$
field, take `the minimal' form with $\pa_\m \rightarrow (\pa_\m -
g{\bf A}_\m \times )$. Identifying ${\bf f} = \d ({\bf F}), \; a=
\d ({\bf A})$ and integrating then yields precisely the full
Yang--Mills equation (17). In the gravitational case, this
argument is strengthened by the fact that the massless spin two
field equations in curved space are in general inconsistent [15];
thus the identification $\phi , \; \g \rightarrow \d (g), \; \d
(\G )$ is the only logical one.

\noindent{\large\bf Appendix B}

\renewcommand{\theequation}{B.\arabic{equation}}
\setcounter{equation}{0}

We describe here the result of the iteration procedure on a
triplet of massless spinless particles, starting from the scalar
representation, rather than the vector one treated in text. The
initial action,
\be%B.1
I_0 = - \int [ \mbox{\boldmath{$\pi$}}^\m \cdot \pa_\m
\mbox{\boldmath{$\phi$}} - \textstyle{\frac{1}{2}}\,
\mbox{\boldmath{$\pi$}}^2] d^4x
 \ee
is invariant under combined isotopic rotations of
 $( \mbox{\boldmath{$\pi$}} ,  \mbox{\boldmath{$\phi$}})$ and also
translations of $\mbox{\boldmath{$\phi$}}$. The isotopic rotations
alone lead to iterated currents and  self-interaction Lagrangians
of the form
\be%B2
j^\m_n = \l^n (\mbox{\boldmath{$\pi$}}^\m \times
\mbox{\boldmath{$\phi$}} ) \times\mbox{\boldmath{$\phi$}} \ldots
\times \mbox{\boldmath{$\phi$}} \; , \;\;\;\; L_n = \l^n
(\mbox{\boldmath{$\pi$}}^\m \times \mbox{\boldmath{$\pi$}} )
\times \mbox{\boldmath{$\phi$}} \ldots \times
\mbox{\boldmath{$\phi$}}_{,\m}
 \ee
Using vector product identities, this sums to
$$%B.3a
I = - \int d^4x [\mbox{\boldmath{$\pi$}}^\m \cdot [{\bf 1} + \l^2
\mbox{\boldmath{$\phi$}}\mbox{\boldmath{$\phi$}} + \times \l
\mbox{\boldmath{$\phi$}}]\cdot \mbox{\boldmath{$\phi$}}_{,\m} (1 +
\l^2 \mbox{\boldmath{$\phi$}}^2)^{-1} -\textstyle{\frac{1}{2}}\,
\mbox{\boldmath{$\pi$}}^2 ] \eqno{(B.3{\rm a})}
$$
$$%B.3b
{\bf j}_\m = \l (1 + \l^2\mbox{\boldmath{$\phi$}}^2)^{-1}[
\mbox{\boldmath{$\pi$}}^\m \times  \mbox{\boldmath{$\phi$}} + \l
(1 + \l^2 )^{-1} \cdot ( \mbox{\boldmath{$\phi$}}
\mbox{\boldmath{$\phi$}} + \mbox{\boldmath{$\phi$}}^2 {\bf
1})\cdot \mbox{\boldmath{$\pi$}}] \; .
 \eqno{(B.3{\rm b})}
$$
\renewcommand{\theequation}{B.\arabic{equation}}
\setcounter{equation}{3}
 To reach the more familiar second-order
form, we use the easily derivable equivalence between the first
and second order actions
 \be%B.4
I = - \int [\mbox{\boldmath{$\phi$}} ^\m \cdot {\bf
P}(\mbox{\boldmath{$\phi$}})\cdot \pa_\m \mbox{\boldmath{$\phi$}}
-\, \textstyle{\frac{1}{2}}\, \mbox{\boldmath{$\pi$}}^2 ]
\leftrightarrow I = -\,\textstyle{\frac{1}{2}}\int
\mbox{\boldmath{$\phi$}}_{,\m}\cdot {\bf
P}^2(\mbox{\boldmath{$\phi$}} ) \cdot
\mbox{\boldmath{$\phi$}}_{,\m}
 \ee
for symmetric dyadics ${\bf P}(\mbox{\boldmath{$\phi$}})$. Then
(B.3) become
$$%B.5a
I = -\textstyle{\frac{1}{2}}\int
\mbox{\boldmath{$\phi$}}_{,\m}\cdot  ({\bf 1} + \l^2
\mbox{\boldmath{$\phi$}}\mbox{\boldmath{$\phi$}}) \cdot
\mbox{\boldmath{$\phi$}}_{,\m} (1 + \l^2
\mbox{\boldmath{$\phi$}}^2 )^{-1} \eqno{(B.5{\rm a})}
$$
$$%B.5b
j_\m = \l \mbox{\boldmath{$\phi$}}_{,\m} \times
\mbox{\boldmath{$\phi$}} (1 + \l^2
\mbox{\boldmath{$\phi$}}^2)^{-1} \; .
 \eqno{(B.5{\rm b})}
$$

The conserved current (B.5b) is just the isotopic part of the
usual chiral current, in the representation in which it reads
 \renewcommand{\theequation}{B.\arabic{equation}}
\setcounter{equation}{5}
 \be%B.6
{\bf j}^c_\m = [ \mbox{\boldmath{$\phi$}}_{,\m} + \l
\mbox{\boldmath{$\phi$}}
 \times \mbox{\boldmath{$\phi$}}_{,\m}] (1 + \l^2
 \mbox{\boldmath{$\phi$}}^2 )^{-1} \; .
 \ee
The $\pa_\m \mbox{\boldmath{$\phi$}}$ part could be obtained by
introducing an initial $\s$ field and iterating on the combined
initial chiral invariance.\footnote{One could also start from {\it
two} triplets in (B.1), which would correspond to the $SU_2 \times
SU_2$ initial Sugawara form, namely a sum of two actions of the
form (24).}

If we consider the combined rotations and translations of (B.1),
with $\d\mbox{\boldmath{$\phi$}} = \mbox{\boldmath{$\phi$}}\times
\mbox{\boldmath{$\o$}} \pm \l^{-1} \mbox{\boldmath{$\o$}}$ (which
are not quite of the chiral form) we get
 \be%B.7
{\bf j}^n_\m = \mbox{\boldmath{$\pi$}}^\m (\times
\l\mbox{\boldmath{$\phi$}} \pm {\bf 1} )^n \; , \;\;\;\; L_n =
{\bf j}^n_\m \cdot \mbox{\boldmath{$\phi$}}_\m \; .
 \ee
Using the identity
 \be%B.8
\sum^\infty_0 (\times \l \mbox{\boldmath{$\phi$}} \pm {\bf 1})^n =
(\times \l \mbox{\boldmath{$\phi$}} \pm {\bf 1})^{-1} = -
 (\times \l \mbox{\boldmath{$\phi$}} + {\bf 1})({\bf 1} + \l^2
 \mbox{\boldmath{$\phi$}}\mbox{\boldmath{$\phi$}}) (1 + \l^2
 \mbox{\boldmath{$\phi$}}^2)^{-1}
  \ee
these may be summed to yield
 \be%B.9
 I = - \int [\pm \, \mbox{\boldmath{$\pi$}}^\m \cdot ({\bf 1} + \l^2
 \mbox{\boldmath{$\phi$}}\mbox{\boldmath{$\phi$}})\cdot
  \mbox{\boldmath{$\phi$}}_\m - \l \mbox{\boldmath{$\pi$}}^\m
  \times \mbox{\boldmath{$\phi$}}\cdot \mbox{\boldmath{$\phi$}}_\m
  ] (1 + \l^2 \mbox{\boldmath{$\phi$}}^2 )^{-1} -
  \textstyle{\frac{1}{2}}\, \mbox{\boldmath{$\pi$}}^2 ] \; .
  \ee
  In second-order form, this reads simply
$$%B.10a
I = -\textstyle{\frac{1}{2}}\int
\mbox{\boldmath{$\phi$}}_{,\m}\cdot  ({\bf 1} + \l^2
\mbox{\boldmath{$\phi$}}\mbox{\boldmath{$\phi$}}) \cdot
\mbox{\boldmath{$\phi$}}_{,\m} (1 + \l^2
\mbox{\boldmath{$\phi$}}^2 )^{-1} \eqno{(B.10{\rm a})}
$$
$$%B.10b
{\bf j}^\m = [\l \mbox{\boldmath{$\phi$}} \times
\mbox{\boldmath{$\phi$}}_{,\m} \pm \mbox{\boldmath{$\phi$}}_{,\m}
\cdot ({\bf 1} + \l^2
\mbox{\boldmath{$\phi$}}\mbox{\boldmath{$\phi$}})](1 + \l^2
\mbox{\boldmath{$\phi$}}^2 )^{-1}
 \eqno{(B.10{\rm b})}
$$
with $I= -\,\frac{1}{2} \int {\bf j}_\m\cdot{\bf j}_\m$.  The
above current differs from the usual chiral one by the
multiplicative factor $({\bf 1} \times \l^2
\mbox{\boldmath{$\phi$}}\mbox{\boldmath{$\phi$}})$, namely the
extra $\mbox{\boldmath{$\phi$}}\mbox{\boldmath{$\phi$}}$ term in
(B.l0b). Likewise, the Lagrangian ${\bf j}^c_\m {\bf j}^c_\m$
differs from (B.l0a) by an extra denominator $(1 +
\l^2\mbox{\boldmath{$\phi$}}^2)^{-1}$ in the
$(\mbox{\boldmath{$\phi$}}\cdot\mbox{\boldmath{$\phi$}}_\m)^2$
term. This model is then a `dynamical  theory of currents'
different (in its current commutators) from the Sugawara  model.

\newpage

\noindent{\large\bf References}

\begin{enumerate}
\item Kraichnan, R.\ (1955). Thesis, Massachusetts Institute of Technology, 1947; and
{\it Physical Review}, {\bf 98}, 1118.
\item
Papapetrou, A.\ (1948). {\it Proceedings of the Royal Irish
Academy}, {\bf 52A}, 11.
\item
Gupta S.N.\ (1952). {\it Proceedings of the Physical Society of
London}, {\bf A65}, 608.
\item
Feynman, R.P.\ (1956). Chapel Hill Conference.
\item
Thirring, W.\ (1959). {\it Fortschritte der Physik}, {\bf 7}, 79.
\item
Halpern L.\ (1963). {\it Bulletin de l'Acad\'{e}mie r.\ de
Belgique. Classe des sciences}, {\bf 49}, 226.
\item
Yang, C.N.\ and Mills, R.L.\ (1954). {\it Physical Review},  {\bf
96}, 191.
\item
Arnowitt, R.\ and Deser, S.\ (1963). {\it Nuclear Physics}, {\bf
49}, 133.
\item
Utiyama, R.\ (1956).  {\it Physical Review}, {\bf 101}, 1597.
\item
Freund, P.G.O.\ and Nambu, Y.\ (1968). {\it Physical Review}, {\bf
174}, 1741.
\item
Sugawara, H.\ (1968).  {\it Physical Review}, {\bf 170}, 1659.
\item
Sommerfield, C.M.\ (1968). {\it Physical Review}, {\bf 176}, 2019.
\item
Deser, S.\ (1969). {\it Physical Review}, {\bf 187}, 1931; (1970)
with L.\ Halpern, J.\ Grav.\ Rel.\ {\bf 1}, 131.
\item
Deser, S.\ and Laurent, B.\ (1968).  {\it Ann.\ Phys.} {\bf 50}, 76.
\item
Aragone, C. and Deser, S.\ (1970).  {\it Nuov. Cim.}  {\bf 3A}, 709; ibid (1980) {\bf 57B}, 33.
\end{enumerate}

\noindent Subsequent amplifications/generalizations of the gravity
derivation may be found
in:\\
Boulware, D.\  and Deser, S.\ (1975) Ann.\ Phys.\ {\bf 89}, 193;
 (1979) Physica {\bf 96A}, 141, with Kay J.; Deser, S. (1987)  Class.\ Quant.\
Grav.\ {\bf 4}, L99; gr-qc/0910.2975, GRG {\bf 42} 641 (2010).

\end{document}